# Jpeg Image Compression Using Discrete Cosine Transform - A Survey

A.M.Raid[1], W.M.Khedr[2], M. A. El-dosuky[1] and Wesam Ahmed[1]

[1] Mansoura University, Faculty of Computer Science and Information System
[2] Zagazig University, Faculty of Science

*ABSTRACT*

*Due to the increasing requirements for transmission of images in computer, mobile environments, the research in the field of image compression has increased significantly. Image compression plays a crucial role in digital image processing, it is also very important for efficient transmission and storage of images. When we compute the number of bits per image resulting from typical sampling rates and quantization methods, we find that Image compression is needed. Therefore development of efficient techniques for image compression has become necessary .This paper is a survey for lossy image compression using Discrete Cosine Transform, it covers JPEG compression algorithm which is used for full-colour still image applications and describes all the components of it.*

## 1. INTRODUCTION

By entering the Digital Age, the world has faced a vast amount of information. Dealing with this vast amount of information can often result in many difficulties. We must store, retrieve, analyze and process Digital information in an efficient way, so as to be put to practical use.

In the past decade many aspects of digital technology have been developed. Specifically in the fields of image acquisition, data storage and bitmap printing. Compressing an image is significantly different than compressing raw binary data. Images have certain statistical properties which can be exploited by encoders specifically designed for them so, the result is less than optimal when using general purpose compression programs to compress images.

One of many techniques under image processing is image compression. Image compression have many applications and plays an important role in efficient transmission and storage of images. The image compression aims at reducing redundancy in image data to store or transmit only a minimal number of samples And from this we can reconstruct a good accession of the original image in accordance with human visual perception. [21][22][23][30].

### 1.1 Principles Behind Compression

Image Compression addresses the problem of reducing the amount of data required to represent the digital image. We can achieve compression by removing of one or more of three basic data redundancies:

(1) Spatial Redundancy or correlation between neighboring pixel.
(2) Due to the correlation between different colour planes or spectral bands, the Spectral redundancy is founded.

(3) Due to properties of the human visual system ,the Psycho-visual redundancy is founded.

We find The spatial and spectral redundancies when certain spatial and spectral patterns between the pixels and the colour components are common to each other and the psycho-visual redundancy produces from the fact that the human eye is insensitive to certain spatial frequencies.

Various techniques can be used to compress the images to reduce their storage sizes as well as using a smaller space. We can use two ways to categorize compression techniques. [13][14]

**-** Lossy Compression System
  Lossy compression techniques is used in images where we can sacrifice some of the finer details in the image to save a little more bandwidth or storage space.

- Lossless compression system
  Lossless Compression System aims at reducing the bit rate of the compressed output without any distortion of the image. The bit-stream after decompression is identical to the original bit-stream.

**-** Predictive coding
  It is a lossless coding method, which means the value for every element in the decoded image and the original image is identical to Differential Pulse Code Modulation (DPCM). [32]

**-** Transform coding
  Transform coding forms an integral part of compression techniques. the reversible linear transform in transform coding aims at mapping the image into a set of coefficients and the resulting coefficients are then quantized and coded. the first attempts is the discrete cosine transform (DCT) domain. [38]

## 1.2 Atypical Image Coder

Three closely connected components form a typical lossy image compression system, they are (a) Source Encoder (b) Quantizer and (c) Entropy Encoder.

### (a) Source Encoder (or Linear Transformer)
It is aimed at decorrelating the input signal by transforming its representation in which the set of data values is sparse, thereby compacting the information content of the signal into smaller number of coefficients. a variety of linear transforms have been developed such as Discrete Cosine Transform (DCT), Discrete wavelet Transform (DWT), Discrete Fourier Transform (DFT).[33][34]

### (b) Quantizer
Aquantizer aims at reducing the number of bits needed to store transformed coefficients by reducing the precision of those values. Quantization performs on each individual coefficient i.e. Scalar Quantization (SQ) or it performs on a group of coefficients together i.e. Vector Quantization (VQ).[35][36]

### (c) Entropy Coding
Entropy encoding removes redundancy by removing repeated bit patterns in the output of the Quantizer. the most common entropy coders are the Huffman Coding, Arithmetic Coding, Run Length Encoding (RLE) and Lempel-Ziv (LZ) algorithm.[34]

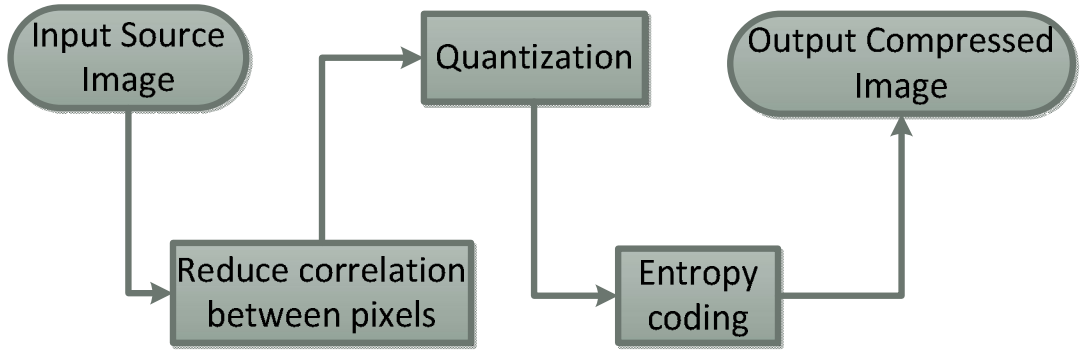

Figure 1.represents the encoding of image compression system

### 1.3 Performance Criteria in Image Compression

We can estimate the performance by applying the following two essential criteria: the compression ratio (CR )and the quality measurement of the reconstructed image( PSNR)

(a) Compression ratio
The Compression ratio (CR) is the ratio between the original image size and the compressed image size.[37]

$$CR = n1/n2 \quad (1)$$

(b) Distortion measure
Mean Square Error (MSE) is a measure of the distortion rate in the reconstructed image.

$$MSE = \frac{1}{HW} \sum_{i=1}^{H} \sum_{j=1}^{W} [X(i,j) - Y(i,j)]^2 \quad (2)$$

- PSNR has been accepted as awidely used quality measurement in the field of image compression.
.

$$PSNR = 10\log_{10} \frac{255^2}{MSE}(dB) \quad (3)$$

## 2. DCT TRANSFORMATION

The most popular technique for image compression, over the past several years, was Discrete cosine transform (DCT). Its selection as the standard for JPEG is One of the major reasons for its popularity. DCT is used by many Non-analytical applications such as image processing and signal-processing DSP applications such as video conferencing. The DCT is used in transformation for data compression. DCT is an orthogonal transform, which has a fixed set of basis function.Dct is used to map an image space into a frequency.[15] DCT has many advantages: (1) It has the ability to pack energy in the lower frequencies for image data. (2) It has the ability to reduce the blocking artefact effect and this effect results from the boundaries between sub-images become visible.[19]

We explain the basics of JPEG compression and decompression in the rest of this paper.

## 3. JPEG COMPRESSION

JPEG Standard is the very well known ISO/ITU-T standard created in the late 1980s. jpeg standard is targeted for full- color still frame applications. one of the most common compression

41

standard is the JPEG standard . Several modes are defined for JPEG including [1][2][3] baseline, lossless, progressive and hierarchical.

The most common mode uses the discrete cosine transform is the JPEG baseline coding system, also it is suitable for most compression applications. Despite being developed for low compressions JPEG it is very helpful for DCT quantization and compression.

JPEG compression reduces file size with minimum image degradation by eliminating the least important information. But it is considered a lossy image compression technique because the final image and the original image are not completely the same and In lossy compression the information that may be lost and missed is affordable. JPEG compression is performed in sequential steps.[11][21][22][23]

- **JPEG Process Steps for color images**

   This section presents jpeg compression steps
    - An RGB to YCbCr color space conversion ( color specification )
    - Original image is divided into blocks of 8 x 8.
    - The pixel values within each block range from[-128 to 127] but pixel values of a black and white image range from [0-255] so, each block is shifted from[0-255] to [-128 to 127].
    - The DCT works from left to right, top to bottom thereby it is applied to each block.
    - Each block is compressed through quantization.
    - Quantized matrix is entropy encoded.
    - Compressed image is reconstructed through reverse process. This process uses the inverse Discrete Cosine Transform (IDCT). [24][26][28]

Figure 2. represents the encoder and decoder block diagrams for colour images.

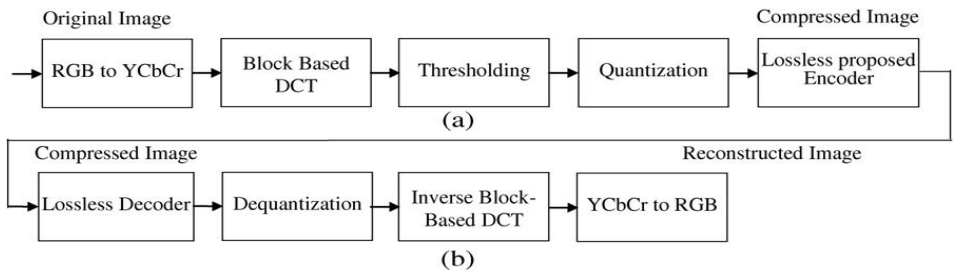

Figure 2. Compression algorithm scheme: (a) compression step and (b) decompression step

## 3.1 Color Specification

The YUV colour coordinate defines Y, Cb, and Cr components of one color image, where Y is commonly called the luminance and Cb, Cr are commonly called the chrominance. the RGB primary uses colour display to mix the luminance and chrominance attributes of a light. Describing of a colour in terms of its luminance and chrominance content separately enable more efficient processing and transmission of colour signals in many applications. To obtain this goal, various three-component colour coordinates have been developed, in which one component(Y) reflects the luminance and the other two collectively (Cb,Cr) characterize hue and saturation. The [Y Cb Cr] T values in the YUV coordinate are related to the [R G B]T values in the RGB coordinate by

$$\begin{pmatrix} Y \\ Cb \\ Cr \end{pmatrix} = \begin{pmatrix} 0.299 & 0.587 & 0.114 \\ -0.169 & -0.334 & 0.500 \\ 0.500 & -0.419 & -0.081 \end{pmatrix} \begin{pmatrix} R \\ G \\ B \end{pmatrix} + \begin{pmatrix} 0 \\ 128 \\ 128 \end{pmatrix} \qquad (4)$$

Similarly, if we want to transform the YUV coordinate back to RGB coordinate, the inverse matrix can be calculated from (4), and the inverse transform is taken to obtain the corresponding RGB components.[4][5][6][7]

### 3.2 Discrete Cosine Transform

After colour coordinate conversion, the next step is to divide the three colour components of the image into many 8×8 blocks. For an 8-bit image, in the original block each element falls in the range [0,255]. Data range that is centred around zero is produced after subtracting The mid-point of the range (the value 128) from each element in the original block, so that the modified range is shifted from [0,255] to [-128,127]. Images are separated into parts of different frequencies by the DCT. The quantization step discards less important frequencies and the decompression step uses the important frequencies to retrieve the image. [27]

This equation gives **the forward 2D_DCT transformation:**

$$F(u,v) = \frac{2}{N} C(u)C(v) \sum_{x=0}^{N-1} \sum_{y=0}^{N-1} f(x,y) \cos\left[\frac{\pi(2x+1)u}{2N}\right] \cos\left[\frac{\pi(2y+1)v}{2N}\right]$$

for $u = 0,...,N-1$ and $v = 0,...,N-1$ \qquad (5)

where $N = 8$ and $C(k) = \begin{cases} 1/\sqrt{2} & \text{for } k = 0 \\ 1 & \text{otherwise} \end{cases}$

This equation gives **the inverse 2D_DCT transformation:**

$$f(x,y) = \frac{2}{N} \sum_{u=0}^{N-1} \sum_{v=0}^{N-1} C(u)C(v) F(u,v) \cos\left[\frac{\pi(2x+1)u}{2N}\right] \cos\left[\frac{\pi(2y+1)v}{2N}\right] \qquad (6)$$

for $x = 0,...,N-1$ and $y = 0,...,N-1$ where $N = 8$

After dct transformation, the "DC coefficient" is the element in the upper most left corresponding to (0,0) and the rest coefficients are called "AC coefficients.[1][4][5][6][7][9][16][17][18][20][29]

### 3.3 Quantization in JPEG

We actually throw away data through the Quantization step. We obtain the Quantization by dividing transformed image DCT matrix by the quantization matrix used . Values of the resultant matrix are then rounded off. The quantized coefficient is defined in (6), and the reverse process can be achieved by the (7).

$$F(u,v)_{Quantization} = round\left(\frac{F(u,v)}{Q(u,v)}\right) \quad (6)$$

$$F(u,v)_{deQ} = F(u,v)_{Quantization} \times Q(u,v) \quad (7)$$

Quantization aims at reducing most of the less important high frequency DCT coefficients to zero, the more zeros the better the image will compress. Lower frequencies are used to reconstruct the image because human eye is more sensitive to them and higher frequencies are discarded. Matrix (8) and (9) defines the Q matrix for luminance and chrominance components [8][12][31].

$$Q_Y = \begin{pmatrix} 16 & 11 & 10 & 16 & 24 & 40 & 51 & 61 \\ 12 & 12 & 14 & 19 & 26 & 58 & 60 & 55 \\ 14 & 13 & 16 & 24 & 40 & 57 & 69 & 56 \\ 14 & 17 & 22 & 29 & 51 & 87 & 80 & 62 \\ 18 & 22 & 37 & 56 & 68 & 109 & 103 & 77 \\ 24 & 35 & 55 & 64 & 81 & 104 & 113 & 92 \\ 49 & 64 & 78 & 87 & 103 & 121 & 120 & 101 \\ 72 & 92 & 95 & 98 & 112 & 100 & 103 & 99 \end{pmatrix} \quad (8)$$

$$Q_C = \begin{pmatrix} 17 & 18 & 24 & 47 & 99 & 99 & 99 & 99 \\ 18 & 21 & 26 & 66 & 99 & 99 & 99 & 99 \\ 24 & 26 & 56 & 99 & 99 & 99 & 99 & 99 \\ 47 & 66 & 99 & 99 & 99 & 99 & 99 & 99 \\ 99 & 99 & 99 & 99 & 99 & 99 & 99 & 99 \\ 99 & 99 & 99 & 99 & 99 & 99 & 99 & 99 \\ 99 & 99 & 99 & 99 & 99 & 99 & 99 & 99 \\ 99 & 99 & 99 & 99 & 99 & 99 & 99 & 99 \end{pmatrix} \quad (9)$$

After quantization, the "zig-zag" sequence orders all of the quantized coefficients as shown in Figure 3 .In the "zig-zag" sequence, firstly it encodes the coefficients with lower frequencies (typically with higher values) and then the higher frequencies (typically zero or almost zero). The result is an extended sequence of similar data bytes, permitting efficient entropy encoding.[10]

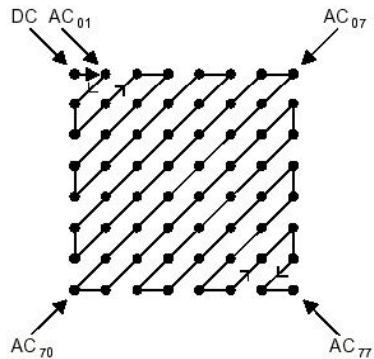

Figure 3. Zigzag Sequencing

### 3.4 Huffman Encoding

Entropy Coding achieves more lossless compression by encoding more compactly the quantized DCT coefficients. Both Huffman coding and arithmetic coding is specified by The JPEG proposal. Huffman coding is used in the baseline sequential codec, but all modes of operation use Huffman coding and arithmetic coding. The source symbols that are not equally probable use Huffman coding efficiently. In 1952 , a variable length encoding algorithm, based on the source symbol probabilities $P(x_i)$, i=1,2……., L is suggested by Huffman . The algorithm achieves the optimality if the average number of bits required to represent the source symbols is a minimum provided the Prefix condition is met. The Huffman algorithm begins with a set of symbols each with its frequency of occurrence (probability) constructing what we can call a frequency table. The Huffman algorithm then builds the Huffman Tree using frequency table. The tree structure contains nodes, each contains a symbol, its frequency, a pointer to a parent node, and pointers to the left and right child nodes. Successive passes through the existing nodes allows the tree to grow. Each pass searches for two nodes that have the two lowest frequency counts, provided that they have not grown a parent node. Anew node is generated when the algorithm finds those two nodes. A new node is assigned as the parent of the two nodes and is given a frequency count that equals the sum of the two child nodes. Those two child nodes are ignored by the next iterations which include the new parent node. The passes stop when only one node with no parent remains. Only one node with no parent will be the root node of the tree. Compression involves traversing the tree beginning at the leaf node for the symbol to be compressed and navigating to the root. The parent of the current node is iteratively selected and seen by this navigation to determine whether the current node is the "right" or "left" child of the parent, therefore determining if the next bit is a (1) or a (0). The final bit string is now to be reversed, because we are proceeding from leaf to root. [25]

### 3.5 Decompression

The compression phase is reversed in the decompression process, and in the opposite order. The first step is restoring the Huffman tables from the image and decompressing the Huffman tokens in the image. Next, the DCT values for each block will be the first things needed to decompress a block. The other 63 values in each block are decompressed by JPEG, filling in the appropriate number of zeros. The last step is combined of decoding the zigzag order and recreating the 8 x 8 blocks .The inverse DCT(IDCT) takes each value in the spatial domain and examines the contributions that each of the 64 frequency values make to that pixel.[7]

## 4. CONCLUSION & FUTURE WORK

Image compression is used for managing images in digital format. This survey paper has been focused on the Fast and efficient lossy coding algorithms JPEG for image Compression/Decompression using Discrete Cosine transform. We also briefly introduced the principles behind the Digital Image compression and various image compression methodologies .and the jpeg process steps including DCT, quantization , entropy encoding.

In the future work we will make a comparison between two techniques of the image compression (Discrete cosine transform and Discrete Wavelet transform).

International Journal of Computer Science & Engineering Survey (IJCSES) Vol.5, No.2, April 2014## REFERENCES

[1]   W.B. Pennebaker and J.L. Mitchell, "JPEG Still Image Data Compression Standards", New York :Van Nostrand Reinhold. (1993).

[2]   Rafael C. Gonzalez, Richard E. Woods, and StevenL. Eddins.''Digital Image Processing Using MATLAB''.. ISBN-10:0130085197. ISBN-13: 978-0130085191. Prentice Hall, 1st edition (September 5, 2003).

[3]   G. K .Wallace:"The JPEG Still Picture Compression Standard," Communication of the ACM, Vol.34, No.4, pp .30-44 ,(Apr 1991).

[4]   Anil K. Jain, ``Fundamentals of digital image processing," Englewood Cliffs: Prentice _Hall information and system sciences series. Prentice _Hall International, London, 1989.

[5]   Liu Chien-Chih and Hang Hsueh-Ming, "Acceleration and Implementation of JPEG 2000 Encoder on TI DSP platform" Image Processing, 2007. ICIP 2007. IEEE International Conference on, Vo1. 3, pp. III-329-339, 2005 .

[6]   ISO/IEC : Information technology-JPEG 2000 image coding system-Part 1: Core coding system. ISO/IEC 15444-1:2000(ISO/IEC JTC/SC 29/WG 1 N1646R ) ( March 2000).

[7]   Hudson, G.P., Yasuda, H., & Sebestyén, I. The international standardization of a still picture compression technique. In Proceedings of the IEEE Global Telecommunications Conference,IEEE Communications Society, pp. 10161021, Nov. 1988.

[8]   John C. Russ.The Image Processing Handbook.,ISBN-10: 0849372542. ISBN-13: 978- 0849372544, CRC Press,5thedition, (December 19, 2006).

[9]   N. Ahmed, T. Natarajan, and K. R. Rao, Discrete Cosine Transform, IEEE. Trans. Computer, Vol C-23, pp. 90-93 Jan 1974.

[10]  Rafael C. Gonzalez and Richard E. Woods. "Digital Image Processing", ISBN-10: 013168728X. ISBN-13: 978-0131687288, Prentice Hall, 3rdedition, (August 21, 2007).

[11]  Boliek, M., Gormish, M. J., Schwartz, E. L., and Keith, A. Next Generation Image Compression and Manipulation Using CREW, Proc. IEEE ICIP, 1997.

[12]  A. B. Watson, "DCT quantization matrices visually optimizedfor individual images," B. Rogowitz and J. Allebach, eds., Human Vision, Visual Processing, and Digital Display IV, SPIE Proc. 1913, (SPIE: Bellingham, WA), pp. 202-216, 1993.

[13]  Shannon, C. E and W. Weaver, The mathematical theory of communication .University of Illinois Press, Urbana,1949.

[14]  A. K Jain, Fundamentals of digital image processing , Prentice Hall , Englewood Cliffs, NJ, 1989.

[15]  Bonnie L.Stephens, Student Thesis on "Image Compression algorithms",California State University, Sacramento, August 1996.

[16]  K. R. Rao and P. Yip, Discrete cosine transform: Algorithms, advantages, applications. San Diego, CA: Academic Press, 1990.

[17]  S. A. Martucci."Symmetric convolution and the discrete sine and cosine transforms. IEEE Transactions Sig.Processing ,SP-42, 1038-1051 (1994).

[18]  E. Feig, S. Winograd. Fast algorithms for the discrete cosine transform.IEEE Transactions on SignalProcessing,vol. 40, no. 9 pp 2174-2193, 1992.

[19]  Randall C. Reiningek and Jerry D. Gibson, "Distributions of the Two-Dimensional DCT Coefficients for Images", IEEE Transactions on Communications, Vol. 31, Issue 6, June 1983.

[20]  Parametric image reconstruction using the discrete cosine transform for optical tomography, Xuejun Gu, Kui Ren and James Masciotti Andreas H. Hielscher, Journal of Biomedical Optics 14!6", 064003 !November/December 2009.

[21]  A. B. Waston Mathematica Journal, vol. 4, no. 1, pp. 81-88, 1994. http://vision.arc.nasa.gov/ Image Compression Using Discrete Cosine Transform, Andrew B. Waston publications/mathjournal94.pdf

[22]  J.J.Ding and J.D.Huang, "Image Compression by Segmentation and Boundary Description", Master's Thesis, National Taiwan University, Taipei, 2007.

[23]  Ken Cabeen and Peter Gent, Image Compression and the Discrete Cosine Transform, Math 45, College of the Redwoods.

[24]  Digital Compression and Coding of Continuous-tone Still Images, Part 1, Requirements and Guidelines.ISO/IEC JTC1 Draft International Standard 10918-1, (Nov. 1991).

[25]  Huffman, D.A. A method for the construction of minimum redundancy codes, Proceedings IRE, vol. 40, pp 1098-1101,1962.
46